\documentclass[11pt,twoside]{article}
\usepackage{graphicx}  
\begin{document}
\begin{center}

\Large {\bf Interstellar extinction by fractal polycrystalline graphite clusters?}

\vspace{1pc}
\large

A.C.\,Andersen,$^{1}$ 
J.A.\,Sotelo,$^{2,4}$  
V.N.\,Pustovit,$^{3,4}$ and
G.A.\,Niklasson,$^{4}$ 

\vspace{1pc}
\small
\it
$^{1}$Dept. of Astronomy \& Space Physics, P.O.Box 515, 751 20 Uppsala, Sweden\\
$^{2}$Dpto. de Fisica, Informatica y Matematicas, Aptdo. 4314, Lima, Peru \\
$^{3}$Inst. of Surface Chemistry, 17 Gen.\ Naumova str., Kiev 03164, Ukraine\\
$^{4}$Dept. of Materials Science, P.O.Box 534, 751 21 Uppsala, Sweden\\
tel: +46 18 471 5982, fax: +46 18 471 5999, e-mail: anja@astro.uu.se
\vspace{1pc}

\bf
Abstract
\rm
\vspace{1pc}
\small

\begin{minipage}[]{5in}
Certain dust particles in space are expected to appear as clusters
of individual grains. The morphology of these clusters could be fractal or 
compact.
To determine how these structural features would affect the interpretation of 
the observed
interstellar extinction peak at $\sim 4.6~\mu$m, we have calculated
the extinction by compact and
fractal polycrystalline graphite clusters consisting of touching identical 
spheres.
We compare three general methods for computing the extinction of the clusters,
namely, a rigorous solution \cite{GA} and two different discrete-dipole
approximation methods -- MarCODES \cite{markel} and DDSCAT \cite{draine+flatau00}.
\end{minipage}
\end{center}
\normalsize

\vspace{1pc}
\section{Introduction}

The shape of many interstellar
grains are expected to be non-spherical and maybe even highly irregular.
One way to deal with irregular particles and clusters of
dust grains is to assume that they consist of touching spheres.
With such an assumption it is possible to construct a variety of
morphologies which can then be compared with observations.

Fitzpatrick \& Massa \cite{fitzpatrick+massa86} studied the interstellar
extinction in the direction of 45 reddened stars, and found that it displays
a peak whose central wavelength $\lambda_{0}$ is remarkably constant
($\lambda_{0} = 2174.4 \pm 17$ {\AA}), even though its full width at half
maximum (FWHM) varies considerably from $~360$ to $~600$ {\AA}; they also
found no apparent correlation between the small variation in $\lambda_{0}$
and the large variation in the FWHM. These characteristics have since been
known as the ``2200~{\AA} peak''.
Graphite is a very promising though controversial
candidate for explaining the 2200 {\AA} peak
\cite{draine88,draine+malhotra,rouleau}.

\section{Method}
We investigate the effect of cluster shape in the extinction of clusters
of spherical polycrystalline graphite particles. Table \ref{clusters}
shows all the clusters we have used in this work; they are either sparse
or compact, and small or large. Their extinction is computed using a 
rigorous solution (GA) \cite{GA} as well as two DDA implementations
-- MarCODES \cite{markel} and DDSCAT 
\cite{draine+flatau00} -- this allows us to test how well the DDA performs 
when applied to clusters of different geometries. 

The three dimensional clusters of 7, 49, 343 ($7^{n}, n=1,2,3$)
particles correspond to the first three stages in the recursive building
of the snowflake fractal.
The dimension of this fractal is $D = {\rm ln 7} / {\rm ln 3} = 1.77$ \cite{vicsek}.
Although such a deterministic structure
is not expected to occur in nature, its fractal dimension is close to that of more
realistic random cluster-cluster aggregation models. In particular, small particles in space
may move in straight,
ballistic trajectories and form larger aggregates upon collisions. Numerical
simulation of this process yields that the fractal dimension of the resulting
aggregates is around 1.9 \cite{meakin+jullien}
and this value has also been confirmed
by experimental results \cite{wurm}.  Moreover, optical properties of 
fractal clusters with
$D<2$ are predicted to be significantly different from those with $D>2$ \cite{berry};
hence we consider it important to use a realistic fractal dimension in our 
computations.

Theoretical computation of absorption and scattering by graphite particles
is difficult because graphite is a semi-metal with high anisotropy 
$(\epsilon_{\parallel},\epsilon_{\perp})$.
In this work we use the dielectric functions $\epsilon_{\perp}$ and 
$\epsilon_{\parallel}$ derived by \cite{draine+lee}, and we deal with 
the anisotropy of graphite by assuming that in
all our clusters, each individual particle is polycrystalline having a
dielectric function $\epsilon_{\rm ave}$ given by the arithmetic average of
$\epsilon_{\parallel}$ and $\epsilon_{\perp}$, namely
$\epsilon_{\rm ave} = \frac{1}{3} \epsilon_{\parallel}+ 
\frac{2}{3} \epsilon_{\perp}$.
By contrast, the usual ``1/3$-$2/3'' approximation, treats individual 
particles as
mono-crystalline - 1/3 of the cluster particles are assumed to have
dielectric function $\epsilon_{\parallel}$ and the remaining 2/3
to have dielectric function $\epsilon_{\perp}$. Considering
grain formation and grain growth in stellar environments \cite{sedlmayr},
polycrystalline particles seems to be a better choice than
mono-crystalline ones.

\begin{table}
\caption{The clusters presented in this paper have three
different geometries: fractal (frac; D$=1.77$), face-center-cubic
(fcc) and simple-cubic (sc).}
\label{clusters}
\footnotesize
\vspace{3pt}
\begin{center}
\begin{tabular}{|c||c|c|c|c|c|c|c|c|c|c|} \hline
Structure & frac & frac & frac & fcc & fcc & fcc & fcc & sc & sc \\
\# particles &  7 & 49 & 343 & 4 & 32 & 49 & 108 & 8 & 27 \\
Designation &  frac7  & frac49& frac343& fcc4& fcc32& fcc49 & fcc108& sc8& sc27 \\ \hline
\end{tabular}
\end{center}
\end{table}

A rigorous and complete solution to the multi-sphere light scattering
problem has been given by G\'erardy \& Ausloos (GA) \cite{GA}.
It is based on 
the exact solution of Maxwell's equations for arbitrary cluster geometries,
polarisation  and incidence direction of the light.
This gives a system of $2NL(L+2)$ equations whose solution is the
$2^L$-polar approximation to the electro-magnetic response of the cluster.
The smallest L needed for the convergence of the extinction will
be different in different spectral regions. In the UV-visible
range it is enough to use L = 7  for open clusters, or L = 9
for compact ones, to compute the extinction with an accuracy of 1\%;
this holds for clusters of up to a few tens of particles.

The discrete dipole approximation (DDA), on the other hand is one of
several discretisation methods for solving scattering
problems in the presence of targets of arbitrary geometry.
In this work we use the DDSCAT code version 5a10 \cite{draine+flatau00}
and  MarCoDES \cite{markel}.

\section{Results}

\begin{figure}[]
\centerline{\includegraphics[width=4.8in]{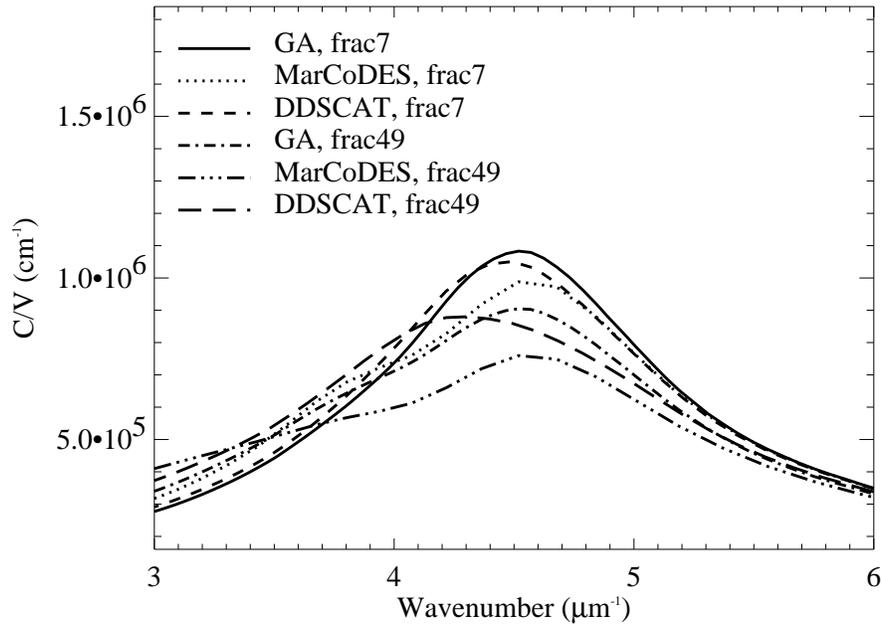}}
\caption{The extinction efficency for the two fractal clusters, frac7 and frac49, 
as calculated with the exact GA solution and the two DDA codes: DDSCAT and MarCoDES.}
\label{results_frac}
\end{figure}

\begin{figure}[]
\centerline{\includegraphics[width=4.8in]{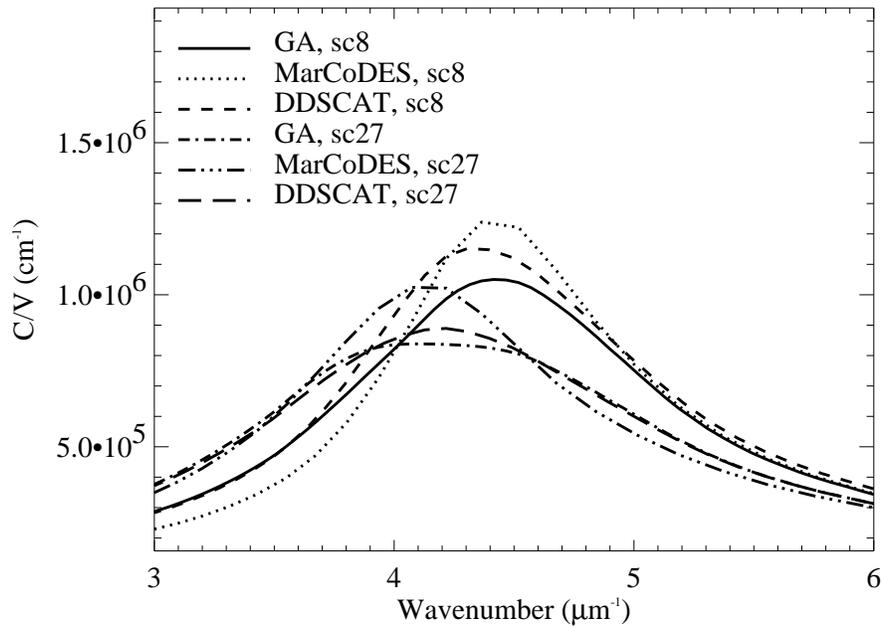}}
\caption{The extinction efficency for the two compact clusters, sc8 and sc27, as
calculated with the exact GA solution and the two DDA codes: DDSCAT and MarCoDES.}
\label{results_sc}
\end{figure}

The results from the three methods for the frac7, frac49, sc8 and sc 27 
cluster can be
seen in Fig.\,\ref{results_frac} and \ref{results_sc}. It is clear that there are significant 
differences between the results of the three codes. MarCODES uses one
dipole per particle, and the difference from the GA results is due to the
neglect of higher order terms, $L\ge 2$. Even in the DDSCAT computations the
number of dipoles used per particle (977 for frac7, 2103 for sc8, 35 for
frac49 and 622 for sc27) was probably not enough to ensure full 
convergence of the result. 

\begin{table}
\caption{Peak position [$\mu$m$^{-1}$] and FWHM [$\mu$m$^{-1}$] of different clusters, as calculated
with the GA method. The value of L indicates at which polar
order the GA calculations were truncated.}
\label{table_ext}
\footnotesize
\vspace{3pt}
\begin{center}
\begin{tabular}{|c||c|c|c|c|c|c|c|c|c|c|c|} \hline
Cluster &frac7 &frac49 &frac343 &sphere &fcc4 &fcc32 &fcc 49 &fcc108 &sc8 &sc27 \\
 L & 11 &6 & 2 & 3 & 11 &  7 & 6 &3 &11 &7 \\
Peak  & 4.52 & 4.52& 4.42& 4.62& 4.43& 3.98& 3.71 & 3.98 & 4.41& 4.12 \\
FWHM  & 1.17 & 1.33 & 1.27 & 0.96 & 1.15 & 1.43 & 1.52& 2.01& 1.27& 1.62 \\ \hline
\end{tabular}
\end{center}
\end{table}

Table \ref{table_ext} lists the extinction's peak position and width for 
all the clusters, as computed with the GA method;
the polar order at which the GA calculation was truncated is
also indicated. The width was determined as the FWHM. For
the asymmetric compact cluster fcc49 we have investigated
the importance of the orientation of the cluster and it was found
that the peak position was shifted less than 0.1 $\mu$m$^{-1}$ while
its width remained the same.
The clusters considered by \cite{rouleau} all
had peak positions at higher wavenumber than the 2200 {\AA} peak
while our clusters all have peak position at lower wavenumbers.
The main difference between our study and the one in
\cite{rouleau} is the way we deal
 with the anisotropy of the graphite grains.  They
\cite{rouleau} used the usual ``$1/3-2/3$'' approximation.
The frac7 and frac49 clusters come close (peak position is
off by $0.04-0.08~\mu$m$^{-1}$) to the observational constrains.
This indicates that small ($N \sim 5 - 100$)
fractal clusters ought to be investigated in more detail
to determine if fractal clusters with low fractal
dimension have a stable peak position around $4.6~\mu$m$^{-1}$
and produce a variable width depending on the number of particles in the cluster.

\subsection*{Acknowledgments}
We would like to thank B.T. Draine and V.A. Markel for making their
      DDA codes available as share-ware. VP would like to thank
       V.A. Markel for fruitful discussions.
      ACA acknowledges support from the Carlsberg Foundation.
      JS and GN acknowledges support from the Swedish Natural Science Research Council (NFR).
      VP acknowledges support from the Wenner-Gren Foundation.


\begin{thebibliography}{9}
   \bibitem{GA} G\'erardy J.M., Ausloos M., 1982, Phys.\ Rev.\ B 25, 4204
   \bibitem{markel} Markel V.A., 1998, User guide for
     MarCoDES - Markel's Coupled Dipole Equation Solvers,
     http://atol.ucsd.edu/~pflatau/scatlib/
   \bibitem{draine+flatau00} Draine B.T., Flatau P.J., 2000, User guide
     for the Discrete Dipole Approximation Code
   DDSCAT (Version 5a10), http://xxx.lanl.gov/abs/astro-ph/0008151v3
   \bibitem{fitzpatrick+massa86} Fitzpatrick E.L., Massa D., 1986, ApJ 307, 286
   \bibitem{draine88} Draine B.T., 1988, ApJ 333, 848
   \bibitem{draine+malhotra} Draine B.T., Malhotra S.K., 1993, ApJ 325, 864
   \bibitem{rouleau} Rouleau F., Henning Th., Stognienko R., 1997, A\&A 322, 633
   \bibitem{vicsek} Vicsek T., 1983, J.\ Phys.\ A 16, L647
   \bibitem{meakin+jullien} Meakin P., Jullien R., 1988, J.\ Chem.\ Phys.\ 89, 246
   \bibitem{wurm} Wurm G., Blum J., 1998, Icarus 132, 125
   \bibitem{berry} Berry M.V., Percival I.C., 1986, Opt.\ Acta 33, 577
   \bibitem{draine+lee} Draine B.T., Lee H.M., 1984, ApJ 285, 89
   \bibitem{sedlmayr} Sedlmayr E., 1994, in Molecules in the Stellar Environment,
     LNP 428, ed.\ U.G. J{\o}rgensen (Springer, Berlin), 163
\end{thebibliography}
\end{document}